\documentclass[aps,superscriptaddress,amsmath,onecolumn,amssymb]{revtex4}

\usepackage{subfigure}
\usepackage{amsmath}
\usepackage{pgfplots}
\pgfplotsset{compat=1.5}
\usepgfplotslibrary{groupplots}

\usepackage{color}
\usepackage{xcolor}
\usepackage{bm}
\usepackage{multirow}
\usepackage{placeins}
\usepackage{soul}
\usepackage{color,xcolor}
\usepackage[colorlinks,linkcolor=red,anchorcolor=blue,citecolor=blue,urlcolor=blue]{hyperref}
\usepackage{cleveref}
\usepackage{amsmath}

\begin{document}

\title{Matrix Product State on a Quantum Computer}

\author{Yong Liu}
\email{liuyong09@nudt.edu.cn}
\thanks{Authors to whom any correspondence should be addressed.}
\affiliation{College of Computer Science and Technology, National University of Defense Technology, Changsha 410073, China}

\author{Guangyao Huang}
\affiliation{College of Computer Science and Technology, National University of Defense Technology, Changsha 410073, China}

\author{Yizhi Wang}
\affiliation{College of Computer Science and Technology, National University of Defense Technology, Changsha 410073, China}

\author{Junjie Wu}
\email{junjiewu@nudt.edu.cn}
\affiliation{College of Computer Science and Technology, National University of Defense Technology, Changsha 410073, China}

\begin{abstract}

Solving quantum many-body systems is one of the most significant regimes where quantum computing applies. Currently, as a hardware-friendly computational paradigms, variational algorithms are often used for finding the ground energy of quantum many-body systems. However, running large-scale variational algorithms is challenging, because of the noise as well as the obstacle of barren plateaus. In this work, we propose the quantum version of matrix product state (qMPS), and develop variational quantum algorithms to prepare it in canonical forms, allowing to run the variational MPS method, which is equivalent to the Density Matrix Renormalization Group method, on near term quantum devices. Compared with widely used methods such as variational quantum eigensolver, this method can greatly reduce the number of qubits required, and thus can mitigate the effects of Barren Plateaus while obtain comparable or even better accuracy. Our method holds promise for distributed quantum computing, offering possibilities for fusion of different computing systems.

\end{abstract}

\date{\today}

\maketitle

\section{Introduction}

Solving quantum many-body systems is a fundamental challenge. Classical computers usually have to deal with the exponential growth in size of the system to mimic the quantum effects. As Feynman raised the concept of quantum simulations~\cite{Feynman1982}, quantum computing begins to show its great power for specific tasks and stimulated the possibility of solving these systems with advantages over the classical computers. Quantum algorithms were proposed for efficient quantum simulations, and are promising for a wide range of applications such as condensed matter physics~\cite{Anderson1984, Sachdev2023}, quantum chemistry~\cite{Cao2019, McArdle2020}, high-energy physics~\cite{Zohar2016, Martinez2016}, materials~\cite{Monroe2021, Maskara2025}, drug discovery~\cite{Wong2023, Santagati2024}, or combinatory optimization tasks~\cite{Moll2018, Harrigan2021}.

As the quantum advantages have been demonstrated on both superconducting and photonic platforms~\cite{Arute2019,Zhu2022,Madsen2022,Zhong2021,Deng2023,Acharya2025,Gao2025}, quantum computing has developed rapidly and is moving towards the noisy intermediate scale quantum (NISQ) era~\cite{Preskill2018}. However, taking the practical physical devices into consideration, running large-scale quantum algorithms still meets severe challenges. For example, algorithms such as the variational quantum eigensolver (VQE)~\cite{Peruzzo2014, Kandala2017}, are troubled with `Barren Plateaus'~\cite{Liu2022, Wang2021}, and are more time consuming to converge to an accurate result. To mitigate these challenges, various distributed quantum frameworks have been proposed. For example, deepVQE can handle large-scale systems with weak correlations among subsystems~\cite{Fujii2022}. Other methods such as circuit cutting~\cite{Peng2020, Tang2022} or hybrid tree tensor networks~\cite{Yuan2021} reduce the sizes of quantum circuits, but requires global optimizations, which limits the efficiency.

Fortunately, the hybrid tensor network provides a general framework for transplanting the classical tensor network method onto quantum devices. Tensor networks are powerful numerical tools for solving fundamental problems in quantum many-body systems. Recently, tensor network methods are applied in quantum information, and become one of the most efficient method for simulating quantum circuits~\cite{Markov2018,Villalonga2019,Villalonga2020,GuoLiu2019,Liu2021,Pan2022}. As one of the most widely used tensor network structures, Matrix Product State (MPS) supports important applications of finding the ground energy of a strongly correlated 1-D many-body system~\cite{Ostlund1995, Hastings2007, Liu2020}. This tensor network structure inspires a quantum circuit ansatz for data encoding or further processing~\cite{Ran2020, Dilip2022, Jaderberg2025}. We focus on a crucial feature of MPS: renormalizing the many-body system and breaking it down into several subtasks executable on quantum devices.

In this work, we propose the idea of {\em quantum Matrix Product State} (qMPS), and develop the {\em variational quantum MPS} (vqMPS) method for quantum simulation, which can be viewed as the quantum version of the Density Matrix Renormalization Group (DMRG) method~\cite{White1992, Schollwock2011} for near-term quantum devices. This method makes it possible to run large-scale quantum simulations with local optimizations on small-scale subsystems, and can reach comparable or even better accuracy over VQE with limited qubits. This can greatly enhance the efficiency for both physical demonstration and numerical simulation. Also, the local optimizations can be distributed on different systems, providing a framework for the fusion of different physical platforms, or further quantum-classical hybrid computing.

\section{The quantum MPS}

A classical MPS contains a group of rank-3 tensors $[A^{\delta_i}]_{\alpha_i, \alpha_{i+1}}$ at the $i$-th site. The left and right dimensions ($\alpha_i$ and $\alpha_{i+1}$) are auxiliary  dimensions representing the correlations between each site. The $N$-qubit state represented by a MPS is
\begin{equation}\label{eq:mps}
  |\psi\rangle = \sum_{\delta_1,\delta_2,\cdots,\delta_N} \mathcal{F}(A^{\delta_1}A^{\delta_2}\cdots A^{\delta_N})|\delta_1\delta_2\cdots\delta_N\rangle,
\end{equation}
where $\mathcal{F}(A^{\delta_1}A^{\delta_2}\cdots A^{\delta_N})$ is the coefficient of basis $|\delta_1\delta_2\cdots\delta_N\rangle$ by contracting the MPS into a complex scalar.

\begin{figure}[t]
  \centering
  \includegraphics[width=\textwidth]{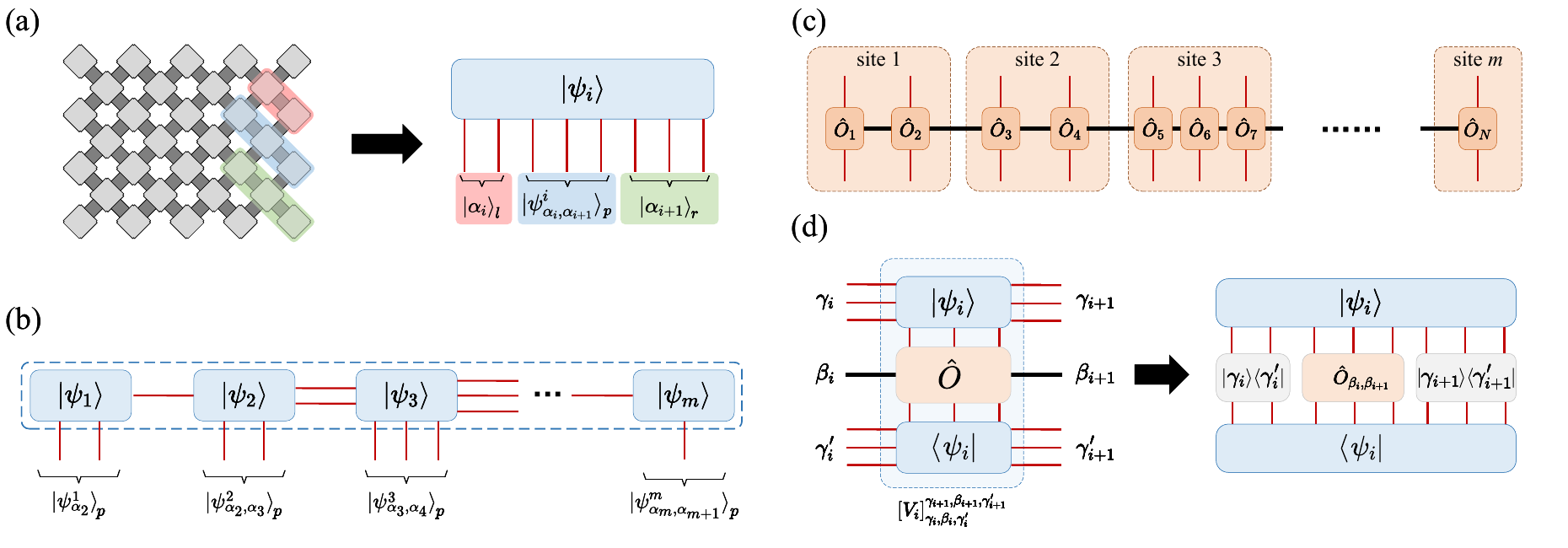}
  \caption{Quantum Matrix Product State model (qMPS). (a) A diagram of implementing a qMPS site on a superconducting processor. A site of qMPS is a quantum state divided into three entangled groups of qubits, labeled by $l$, $p$ and $r$. Qubits in groups $l$ and $r$ are used as auxiliary dimensions of the qMPS, and the qubits in group $p$ are used ad the physical dimensions of the qMPS. (b) The global state represented by a qMPS is determined by contracting all quantum dimensions, as represented by Eq.~\ref{eq:qmps_full_state}. The number of qubits contained in the global states is the sum of qubits in physical dimension of each site. (c) The MPO sites can be divided into $m$ groups to match the number of physical qubits in each qMPS site. (d) The local expectation is represented by a rank-6 classical tensor, with each element calculated through Eq.~\ref{eq:expectation}.}\label{fig:qmps}
\end{figure}

A qMPS is a multi-component structure that represents a large global state through a group of $m$ distributed smaller-scale quantum states. The $i$-th site of the qMPS is a physical quantum state containing $n_i$ qubits, which is given by
\begin{equation}\label{eq:mps_site}
  |\psi_i\rangle = \sum_{\alpha_i,\alpha_{i+1}}c^i_{\alpha_i,\alpha_{i+1}}|\alpha_i\rangle_{l}|\psi_{\alpha_{i},\alpha_{i+1}}^i\rangle_{p}|\alpha_{i+1}\rangle_{r},
\end{equation}
where qubits are partitioned into three entangled groups labeled $l$ (left auxiliary), $p$ (physical) and $r$ (right auxiliary). Fig.~\ref{fig:qmps}(a) is a diagram of implementing a qMPS site on a superconducting processor. A qMPS site is naturally a rank-$n_i$ quantum tensor~\cite{Yuan2021}, and the elements of this tensor are the amplitudes of this state. For consistency with a MPS with a bond dimension of $\chi$, we allocate $n_l = n_r = \lceil\log_2(\chi)\rceil$ qubits per auxiliary dimension, and there can be totally $2\lceil\log_2(\chi)\rceil + n_p$ qubits where $n_p$ is the number of physical qubits used in a qMPS site. Generally $n_p \geq 1$, and we focus on cases where $n_p = 1$ to maintain consistency with the classical cases. The data in this site consist of its amplitudes and can be indexed by projecting the state onto the corresponding basis. The space complexity of qMPS naturally scales logarithmically compared to classical MPS.

Contracting two qMPS sites follows the rules~\cite{Yuan2021} of
\begin{equation}\label{eq:qmps_two_site}
\begin{aligned}
  |\psi_{i, i+1}\rangle =&\sum_{\alpha_{i+1}}\langle \alpha_{i+1}|_{r}\cdot|\psi_i\rangle \otimes \langle \alpha_{i+1}|_{l}\cdot|\psi_{i+1}\rangle\\
=&\sum_{\alpha_{i+1}}\left(\sum_{\alpha_i, \alpha_{i+2}}c^i_{\alpha_i,\alpha_{i+1}}c^{i+1}_{\alpha_{i+1},\alpha_{i+2}}|\alpha_i\rangle_{l}|\psi_{\alpha_{i},\alpha_{i+1}}^i\rangle_{p}|\psi_{\alpha_{i},\alpha_{i+1}}^i\rangle_{p}|\alpha_{i+1}\rangle_{r}\right),
\end{aligned}
\end{equation}
These qMPS sites collectively form a global $N$-qubit state, as represented by equation:
\begin{equation}\label{eq:qmps_full_state}
  |\Psi\rangle = \sum_{\alpha_1, \alpha_2, ..., \alpha_n} \bigotimes_{i = 0}^m (\langle\alpha_i|_{l}\langle \alpha_{i+1}|_{r})\cdot|\psi_i\rangle,
\end{equation}
also as shown in Fig.~\ref{fig:qmps}(b). The number of qubits in the global state is the sum of the number of qubits representing the physical dimension of each qMPS site. The auxiliary dimension, if not bounded, grows exponentially within a classical MPS, while the number of auxiliary qubits grows linearly. Therefore, the qMPS model can be used for strongly entangled systems.

Calculating the global expectation of a qMPS for a global observable can be done through local measurements. In the language of MPS, the observable can first be decomposed into a matrix product operator (MPO), and the corresponding physical dimensions are divided into $m$ groups in correspondence to every qMPS site. For any $N$-qubit observable, it can be written as the sum of Pauli products $\hat{O} = \sum_{i=1}^K c_i (\otimes_{j=1}^N P_{ij})$ where $P_{ij} \in \{I, \sigma_x, \sigma_y, \sigma_z\}$, and further in a simplified and diagonalized MPO form
\begin{equation}\label{eq:MPO}
  \hat{O} = [\hat{O}_1]\times [\hat{O}_2] \times \cdots \times [\hat{O}_N]
  = \begin{bmatrix}P_{11}& \cdots& P_{1K}\end{bmatrix}\times \begin{bmatrix}
  P_{21} & & \\
  & \ddots & \\
  & & P_{2K} \\
  \end{bmatrix}\times\cdots\times\begin{bmatrix}c_1P_{N1}\\ \vdots\\ c_KP_{NK}\end{bmatrix}.
\end{equation}
Note that the multiplications between Pauli operators are tensor products. In practice, it is sufficient to record the appearance of each Pauli operator instead of storing a huge amount of MPO data. For simplicity, the coefficients for each term are concentrated in $\hat{O}_N$. These $N$ MPO sites can be divided into groups to match the number of physical qubits in each qMPS site, as shown in Fig.~\ref{fig:qmps}(c), while in this work, we focus on the cases where $m=K$, i.e., the MPO sites are separated per physical qubit of the qMPS sites.

The calculation of expectation can be done site-by-site and locally on a smaller quantum device, as shown in Fig.~\ref{fig:qmps}(d). Suppose we are now at site $i$ of the qMPS, which is in state $|\psi_i\rangle$. The corresponding MPO site is represented by $[\hat{O}^{\sigma_i, \sigma'_i}]_{\beta_i,\beta_{i+1}}$ where $\sigma_i$ and $\sigma'_i$ are the physical dimensions, while $\beta_i$ and $\beta_{i+1}$ are assistant dimensions linking to the neighbouring sites. As shown in Fig.~\ref{fig:qmps}(d), the MPO site only covers the physical dimensions of $|\psi_i\rangle$. The auxiliary qubits act as free indices, and can be transferred into classical indices by projecting the auxiliary qubits onto the computational basis, obtaining a list of (unnormalized) quantum states as
\begin{equation}
\begin{aligned}
|\psi_i(j, k)\rangle&=\langle j|_l\langle k|_r \cdot|\psi_i\rangle\\
&=\langle j|_l\langle k|_r \left( \sum_{\alpha_i,\alpha_{i+1}}c^i_{\alpha_i,\alpha_{i+1}}|\alpha_i\rangle_{l}|\psi_{\alpha_{i},\alpha_{i+1}}^i\rangle_{p}|\alpha_{i+1}\rangle_{r}\right)\\
&=c_{j, k}^i|\psi_{j, k}^i\rangle.
\end{aligned}
\end{equation}
For an MPO generated following Eq.~\ref{eq:MPO}, it satisfies that $[\hat{O}^{\sigma_i, \sigma'_i}]_{\beta_i, \beta_{i+1}} = \delta_{\beta_i, \beta_{i+1}} P_{i,\beta_i}$, which indicates that $[\hat{O}^{\sigma_i, \sigma'_i}]_{\beta_i, \beta_{i+1}} = {\bf 0}$ given $\beta_i\neq \beta_{i+1}$ and we can only iterate over the pauli operators appearing in its diagonal. Therefore, it also provides two free indices, and as a result, the local expectation is a rank-6 tensor (or a rank-3 tensor if the site is at the end of the chain), as shown in eq.~\ref{eq:expectation}.

\begin{equation}\label{eq:expectation}
\begin{array}{ll}
   [V_i]_{\gamma_i, \beta_i, \gamma'_i}^{\gamma_{i+1}, \beta_{i+1}, \gamma'_{i+1}}&=\langle\psi_i(\gamma'_i, \gamma'_{i+1})|\hat{O}_{\beta_i, \beta_{i+1}}|\psi_i(\gamma_i, \gamma_{i+1})\rangle\\
   &=\left(\langle \psi_i |\cdot|\gamma'_i\rangle_l|\gamma'_{i+1}\rangle_r\right)\hat{O}_{\beta_i, \beta_{i+1}}\left(\langle \gamma_i|_l\langle \gamma_{i+1}|_r\cdot|\psi_i\rangle\right)\\
   &=\langle\psi_i| \left(|\gamma'_i\rangle_{l}\langle\gamma_i|_{l} \otimes \hat{O}_{\beta_i,\beta_{i+1}}\otimes |\gamma'_{i+1}\rangle_{r}\langle\gamma_{i+1}|_{r} \right)|\psi_i\rangle.
\end{array}
\end{equation}

In practice, $\hat{O}_{\beta_i,\beta_{i+1}}$ is a legal observable, and each $|\gamma_i\rangle_{l}\langle\gamma'_i|_{l}$ or $|\gamma_{i+1}\rangle_{r}\langle\gamma'_{i+1}|_{r}$ can be expanded into series of Pauli terms. Then, the local expectation can be measured experimentally. Running over $I$, $\sigma_x$, $\sigma_y$ and $\sigma_z$ for each auxiliary qubit leads to a time complexity of $O(\chi^4)$ for a single term. The global expectation value can be calculated by classical tensor contraction, and the overall complexity to evaluate the global expectation is $O(Km\chi^4)$, where $K$ is the number of terms in the global observable, and $m$ is the number of qMPS sites. Once all local expectations are measured, the global expectation can be calculated through classical tensor contraction
\begin{equation}
V=\sum_{\begin{subarray}{c}
\gamma_1, \gamma_2, ...,\gamma_{m+1}\\
\beta_1, \beta_2, ...,\beta_{m+1}\\
\gamma'_1, \gamma'_2, ...,\gamma'_{m+1}
\end{subarray}}
[V_1]_{\gamma_1, \beta_1, \gamma'_1}^{\gamma_{2}, \beta_{2}, \gamma'_{2}}
[V_2]_{\gamma_2, \beta_2, \gamma'_2}^{\gamma_{3}, \beta_{3}, \gamma'_{3}}
\cdots
[V_m]_{\gamma_m, \beta_m, \gamma'_m}^{\gamma_{m+1}, \beta_{m+1}, \gamma'_{m+1}}.
\end{equation}

\section{Turning a qMPS into canonical form}

\begin{figure*}[!htb]
  \centering
  \includegraphics[width=\textwidth]{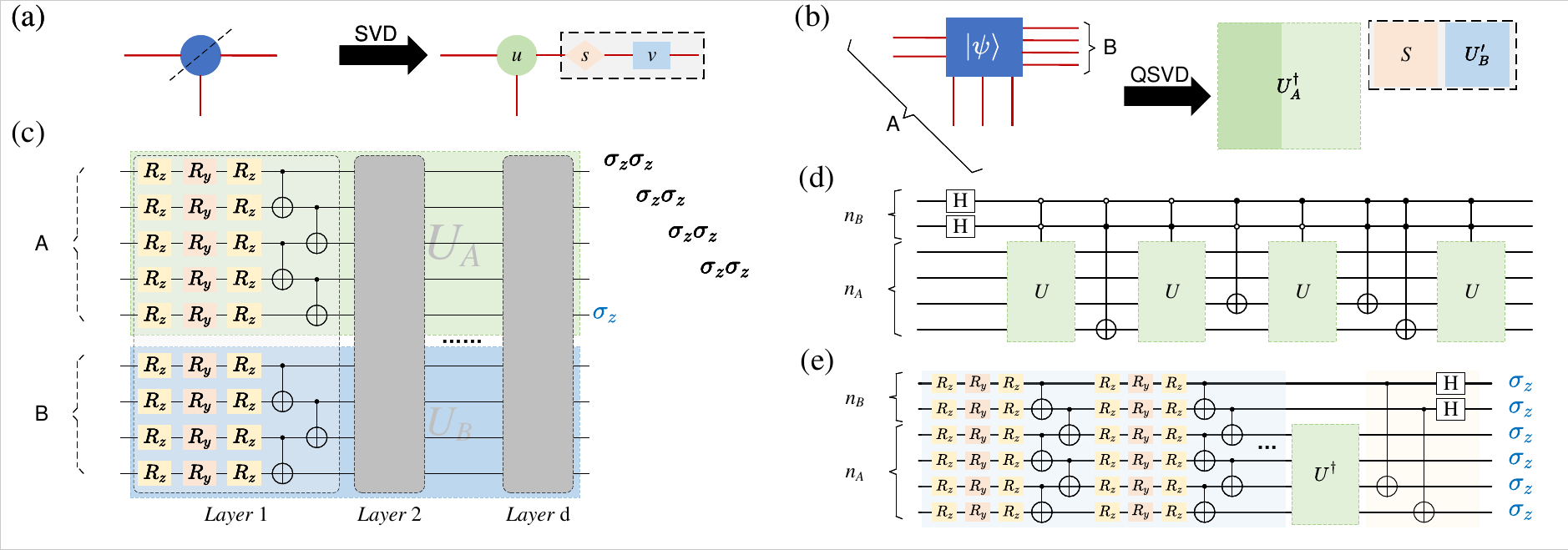}
  \caption{Turning a qMPS site into left-canonical form. (a) Diagram of preparing a classical MPS site into left-canonical form using SVD algorithm. (b) Schematic view of preparing a qMPS site into left-canonical form with a QSVD lagorithm. (c) The variational quantum circuits for finding two unitary matrices to diagonalize the state. The expectation value of observables include $\sigma_z\sigma_z$ on qubits lined with dashed lines, and $\sigma_z$ on the rest qubits. (d) Quantum circuit for quantum reshape operation, transferring the data of a unitary into a quantum state, namely a new qMPS site. It turns the elements of $U$ into the amplitudes of a quantum state $|\Psi_{\rm rs}\rangle = \frac{1}{\sqrt{2^{n_{B}}}}\sum_{i=0}^{2^{n_B} - 1}|i\rangle (U|i\rangle)$. (e) Variational algorithm for quantum reshape. The idea for reshape is to generate the state $|\Psi_{\rm rs}\rangle$ through a variational circuits, and then turn it into maximally entangled state by $(I\otimes U^\dagger)|\Psi_{\rm rs}\rangle=\sum_{i=0}^{2^{n_B} - 1}|i\rangle |i\rangle$.}\label{fig:qsvd}
\end{figure*}

An important step before applying MPS is to turn it into canonical form. For example, Eq.~\ref{eq:canonical} shows how a MPS site $[A^{\delta_i}]_{\alpha_i, \alpha_{i+1}}$ is in right(left)-canonical form.
\begin{equation}\label{eq:canonical}
\begin{aligned}
  \sum_{\delta_i, \alpha_{i+1}} [A^{\delta_i}]_{\alpha_i, \alpha_{i+1}} {\rm conj}\left([A^{\delta_i}]_{\alpha'_i, \alpha_{i+1}}\right) &= \delta_{\alpha_i,\alpha'_i},\\
  \sum_{\delta_i, \alpha_{i}} [A^{\delta_i}]_{\alpha_i, \alpha_{i+1}} {\rm conj}\left([A^{\delta_i}]_{\alpha_i, \alpha'_{i+1}}\right) &= \delta_{\alpha_{i+1},\alpha'_{i+1}},
\end{aligned}
\end{equation}
where ${\rm conj}(A)$ indicates to take the elementwise conjugate of the tensor $A$. For classical MPS, this is done by Singular value decomposition (SVD). Unlike ref.~\cite{Schuhmacher2024} that uses classical ancillary data, here we apply quantum algorhms to turn a qMPS site into canonical form, through quantum singular value decomposition (QSVD) and a quantum reshape algorithm, as shown in Fig.~\ref{fig:qsvd}.

\subsection{Quantum SVD for imbalanced system}
The system is divided into two subsystems, and the main idea for decomposing a state $|\psi\rangle$ is to find two unitaries $U_A$ and $U_B$ for diagonalization, such that $U_A\otimes U_B|\psi\rangle=\sum_i s_i|i\rangle|i\rangle$, as shown in Fig.~\ref{fig:qsvd}(b)~\cite{Carlos2020}. This is done through a variational algorithm optimizing $L=\sum_q (1-\langle\sigma_z^{q,A} \sigma_z^{q,B})/2$, where $\sigma_z^{q,A}$ indicates the measurement on the $q$-th qubit of subsystem A. We note that another variational method exists for decomposing a unitary matrix~\cite{Wang2021}, but in qMPS, it has to deal with a state rather than a matrix, and we have to extend the method for imbalanced systems where the number of qubits in the subsystems is unequal. Without loss of generality, we assume $n_A > n_B$, and the goal of the decomposition is to find the corresponding unitary such that
\begin{equation}
U_A\otimes U_B|\psi\rangle=\sum_{i=0}^{2^{n_B}-1} s_i\left(|i\rangle\otimes|0\rangle^{\otimes{(n_A-n_B)}}\right)\otimes |i\rangle.
\end{equation}
where $|i\rangle$ is an $n_B$-qubit state. The loss function must also account for the remaining qubits, which can be updated as
\begin{equation}\label{eq:svd}
  L_{\rm SVD}=\sum_{q=1}^{n_B} \frac{1-\langle\sigma_z^{q,A} \sigma_z^{q,B}\rangle}{2} - \sum_{q=n_B+1}^{n_A}(1- \langle\sigma_z^{q,A} \rangle),
\end{equation}

We construct a variational circuit for this purpose as shown in Fig.~\ref{fig:qsvd}(c). The circuit is divided into two parts, containing $n_A$ and $n_B$ qubits respectively. In each layer of the circuit, it comprises single-qubit rotations and interlacing CNOT gates. The decomposition requires that if subsystem $A$ is measured as $|i\rangle$ (in the binary representation of its $n_B$ qubits together with the latter $n_A - n_B$ qubits in state $|0\rangle$), then subsystem $B$ should also be in state $|i\rangle$ (in the binary representation of its $n_B$ qubits). This is done through the correlate measurement on the overlapped qubits of the two subsystems through the observable of $\sigma_z\otimes\sigma_z$, meanwhile the rest qubits are measured in $\sigma_z$ basis. Optimization is performed using standard variational algorithms, by gradient descent method. The two unitaries are obtained when the loss is optimized to zero, yielding the decomposition as
\begin{equation}\label{eq:decomposition}
\begin{aligned}
  |\psi\rangle &= \sum_i s_i U_A^\dagger|i\rangle \otimes U_B^\dagger |i\rangle\\
               &\stackrel{\rm reshape}{\longrightarrow} U_A^\dagger S U'_B.
\end{aligned}
\end{equation}
where $S={\rm diag}(s_1, s_2, ...,s_{2^{\min(n_A, n_B)}})$ and $U'_B$ is the conjugate of $U_B$. As shown in Ref.~\cite{Carlos2020}, the loss decays exponentially fast when the depth of the circuit grows. The feasibility of qSVD can be further enhanced by using a more effective ansatz or increasing parametric layers of circuit.

\subsection{Quantum reshape}
To convert a qMPS site into canonical form, one should transfer the result of QSVD (i.e., $U_A^\dagger$ or $U'_B$) into a qMPS site (i.e., a quantum state), and apply the other unitary as well as the diagonal matrix of singular values to the neighbouring site. For the first step, we employ a quantum reshape algorithm. The corresponding circuit to implement this is straight forward, as shown in Fig.~\ref{fig:qsvd}(d). To prepare a left-canonical site, we directly apply $U_A^\dagger$ in this circuit, and it will concatenate the columns of $U_A^\dagger$ into a state vector
\begin{equation}\label{eq:reshape_state}
|\Psi_{\rm rs}\rangle = \frac{1}{\sqrt{2^{n_B}}}\sum_{i = 0}^{2^{n_B}} |i\rangle (U_A^\dagger|i\rangle)
\end{equation}
However, this circuit may be too deep for near-term devices, and we also develop a variational algorithm for this reshape, as shown in Fig.~\ref{fig:qsvd}(e). The idea for this is to approximate $|\Psi_{\rm rs}\rangle$ with a variational circuit. Applying $U_A$ to the latter part of the state produces a maximally entangled state, and can further be transformed into $|0\rangle^{\otimes n}$ using merely CNOT and Hadamard gates. The corresponding cost function is
\begin{equation}\label{eq:svd}
  L_{\rm reshape}=\sum_{q=1}^{n}(1- \langle\sigma_z^q \rangle).
\end{equation}
If $L_{\rm reshape}$ is optimized to 0, the output state can be determined to be $|0\rangle^{\otimes n}$ and the variational circuit produces $|\Psi_{\rm rs}\rangle$ shown in eq.~\ref{eq:reshape_state} ideally. In our numerical simulations, we find that the variational reshape can achieve fidelities exceeding $99\%$ for circuits with no more than 7 qubits. This empirical result is sufficient for our further application. To prepare a right-canonical site, the corresponding data (the rows of $U'_B$) can be obtained similarly.

The latter step is to keep the global state invariant by merging the rest result of QSVD (that is $SU'_B$ or $U_A^\dagger S$) to the neighbouring site. We can also apply another variational algorithm to find the result of applying this operation on the neighbouring site~\cite{Motta2020}, but fortunately, in some cases this step is unnecessary, such as in the variational qMPS (vqMPS) method we will introduce in the next section.

\section{The variational MPS method on a quantum computer}

\begin{figure}[!h]
  \centering
  \includegraphics[width=0.6\textwidth]{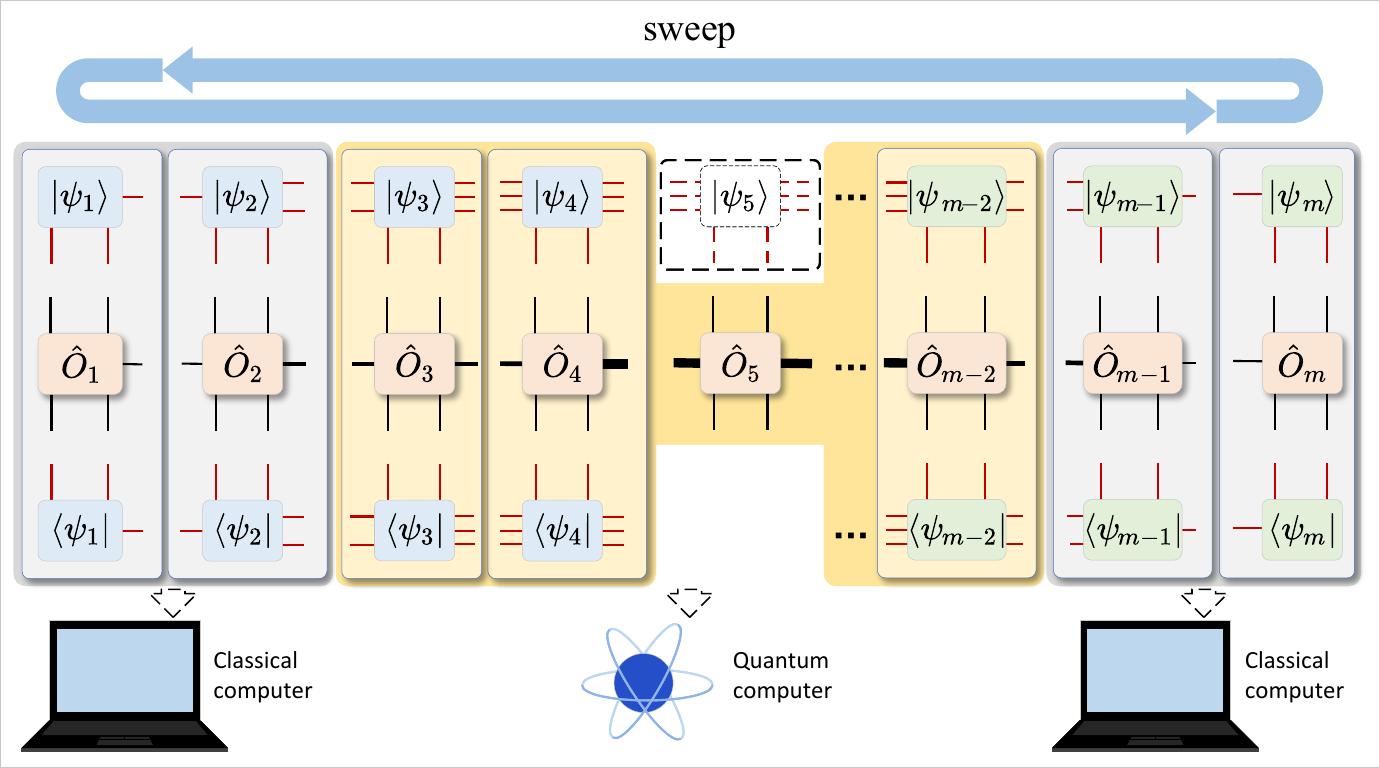}
  \caption{The schematic view of vqMPS method. The process is similar to the classical DMRG method, by sweeping back-and-forth. The global qMPS is initially prepared in left-canonical form (the blue sites), and updated site-by-site from right to left. After this sweeping, the qMPS would be converted into right-canonical (the green sites), and can be updated by sweeping from left to right. The ground energy is estimated during these sweeps. The vqMPS facilitates quantum-classical hybrid computing. The sites on both sides can be updated on classical computers, while updating strongly correlated sites can be accelerated using quantum computing.}\label{fig:vqMPS}
\end{figure}

Perhaps the most important application of MPS is the variational MPS (vMPS) method for finding ground energy of many-body system, which is equivalent to the DMRG method~\cite{White1992,Schollwock2005}. With qMPS and all the components mentioned above, we can run this algorithm on quantum devices, and we denote this method by variatioanl quantum MPS (vqMPS) method. The vMPS algorithm updates the tensors site-by-site (or more than one site at a time) instead of optimizing the whole system. For site $i$, the other qMPS sites and the Hamiltonian MPO constitute its environment, or more specifically the effective Hamiltonian. This site then can be updated by the ground state of the effective Hamiltonian (see Ref.~\cite{Schollwock2005,Schollwock2011} for detials). The vqMPS procedure closely resembles classical vMPS, by performing back-and-forth sweeps, as shown in Fig.~\ref{fig:vqMPS}.

The global qMPS is prepared randomly with $m$ separate quantum states, and turned into left-canonical form for sweeping from right to left. For each site, the number of qubits is determined according to the size of bond dimensions. During a sweeping process, the states are updated site-by-site, through finding the ground state of the effective Hamiltonian.

When updating site $i$ during a sweep from right to left, the effective Hamiltonian $H_{\rm eff}^i$ is constructed by contracting the local tensors obtained following Eq.~\ref{eq:expectation}, as shown in Eq.~\ref{eq:Heff}.
\begin{equation}\label{eq:Heff}
  [H_{\rm eff}^i]_{\gamma'_i, \sigma'_i, \gamma'_{i+1}}^{\gamma_i, \sigma_i, \gamma_{i+1}} = \sum_{\beta_{i}, \beta_{i+1}}
[L_i]^{\gamma_i, \beta_i, \gamma'_{i}}
[\hat{O}^{\sigma_i, \sigma'_i}]_{\beta_i, \beta_{i+1}}
[R_i]_{\gamma_{i+1}, \beta_{i+1}, \gamma'_{i+1}}.
\end{equation}
where the $L$ tensor and $R$ tensor are
\begin{equation}\label{eq:LR}
\begin{aligned}
[L_i]^{\gamma_i, \beta_i, \gamma'_{i}}&=\sum_{\begin{subarray}{c}
\gamma_1, \gamma_2, ..., \gamma_{i-1}\\
\beta_1, \beta_2, ...,\beta_{i-1}\\
\gamma'_1, \gamma'_2, ...,\gamma'_{i-1}
\end{subarray}}
[V_1]_{\gamma_1, \beta_1, \gamma'_1}^{\gamma_{2}, \beta_{2}, \gamma'_{2}}
[V_2]_{\gamma_2, \beta_2, \gamma'_2}^{\gamma_{3}, \beta_{3}, \gamma'_{3}}
\cdots
[V_{i-1}]_{\gamma_{i-1}, \beta_{i-1}, \gamma'_{i-1}}^{\gamma_{i}, \beta_{i}, \gamma'_{i}},\\
[R_i]_{\gamma_{i+1}, \beta_{i+1}, \gamma'_{i+1}}&=\sum_{\begin{subarray}{c}
\gamma_{i+2}, \gamma_{i+3}, ..., \gamma_{m+1}\\
\beta_{i+2}, \beta_{i+3}, ..., \beta_{m+1}\\
\gamma'_{i+2}, \gamma'_{i+3}, ..., \gamma'_{m+1}
\end{subarray}}
[V_{i+1}]_{\gamma_{i+1}, \beta_{i+1}, \gamma'_{i+1}}^{\gamma_{i+2}, \beta_{i+2}, \gamma'_{i+2}}
[V_{i+2}]_{\gamma_{i+2}, \beta_{i+2}, \gamma'_{i+2}}^{\gamma_{i+3}, \beta_{i+3}, \gamma'_{i+3}}
\cdots
[V_m]_{\gamma_m, \beta_m, \gamma'_m}^{\gamma_{m+1}, \beta_{m+1}, \gamma'_{m+1}}.
\end{aligned}
\end{equation}
Note that the indices are reordered, and the effective Hamiltonian $H_{\rm eff}^i$ is then reshaped into matrix form. Then, we find the ground state of this effective Hamiltonian, and decompose it as $U_A^\dagger S U'_B$ through QSVD. This site can then be updated through reshaped $U'_B$. Since we are going to update the $(i-1)$-th site, $U_A^\dagger S$ is no longer necessary to merge.

The ground energy of the whole system can be obtained through the sweeps resembling the classical DMRG, and our method can be scaled up to large-scale systems. For the optimization of each site, the effective Hamiltonian $H_{\rm eff}^i$ can be much smaller than the global Hamiltonian, and thus may mitigate the Barren Plateau by limiting the problem size. For a many-body system containing $N$ qubits, it can be represented by $N$ qMPS sites, with each site containing $2n_\chi + 1$ qubits, and allows various means to find its ground state, such as VQE or perhaps the quantum phase estimation algorithm in the future which can bring quantum advantages. We apply VQE for the effective Hamiltonian because the main target for comparison is the full-scale VQE, using the same circuit ansatz and optimization configurations. It's worth noting that the full-scale VQE optimizes the whole system, and is supposed to obtain better accuracy than optimizing the system site-by-site. However, it suffers from the BP phenomenon when the system reaches a certain scale. The benefit of the divide-and-conquer method is thus it can support large-scale system on limited number of qubits, while the classical DMRG theory allows to reach comparable accuracy for appropriate systems.

In practice, vqMPS can adopt a hybrid computing paradigm: Sites near boundaries with low entanglement can be efficiently updated using classical DMRG on classical computers, while central sites exhibiting strong entanglement reside on quantum devices. Consequently, this strategy leverages the advantages of both classical and quantum computing and avoids frequent quantum-classical interactions appearing in conventional VQE. A variant of VQE that applies a divide-and-conquer scheme is deepVQE~\cite{Fujii2022}, which however is suitable for systems with weak correlation between subsystems. The original hybrid tensor network scheme optimizes the whole system simultaneously~\cite{Yuan2021}, meaning that the optimization of each tensor is correlated with others. In our method, the optimization of each site is relatively independent once the effective Hamiltonian is given, making it more flexible.

\section{Numerical simulation}

We conduct the numerical simulation to demonstrate vqMPS on a personal computer to show evidence for quantum devices to run tensor network algorithms.

\begin{figure}[hbt!]
  \centering
  \includegraphics[width=0.4\textwidth]{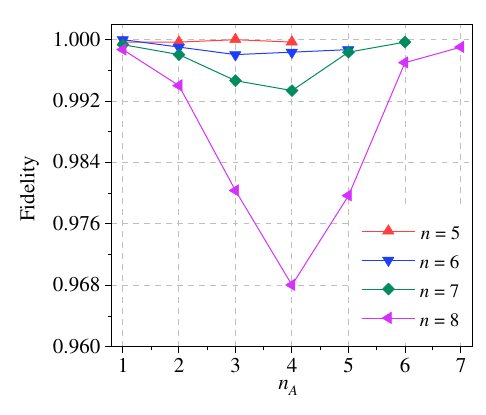}
  \caption{Results of quantum singular value decomposition. For $n$ qubits, we test to decomposing it to $n_A + n_B$ qubits through variational algorithms, rebuild it through the obtained singular values and unitaries, and then measure the fidelity. The overall fidelity recovered reaches over 96\% and it achieves better accuracy when the system is imbalanced.}\label{fig:res_qsvd}
\end{figure}

We first demonstrate the quantum SVD method for imbalanced systems. As a proof of principle, Fig.~\ref{fig:res_qsvd} shows the numerical results of decomposing random states. The ansatz of the variational circuits used are shown in Fig.~\ref{fig:qsvd}(c), but we use only one $R_y$ gate on each qubit in each variational layer for the efficiency of simulation. The result of the decomposition can be validated through the fidelity of the recovered state with the original state. We test to decompose states containing $n=3$ to $8$ qubits. For each case, we first decompose it to $n=n_A + n_B$ qubits using a variational circuit with sufficient depth, then calculate the fidelity between the original state and the state recovered from the learned parameters. Each point is averaged through 3 tests. The recovered state can reach identity fidelities for $n = 3$ and $4$, and the results for more qubits are shown in Fig.~\ref{fig:res_qsvd}. All simulated results reach a fidelity exceeding 96\%, with fidelities above 99\% for systems involving up to 7 qubits. The fidelity can be further improved by increasing the depth of the variational circuits.

\begin{figure*}[t!]
  \centering
  \includegraphics[width=\textwidth]{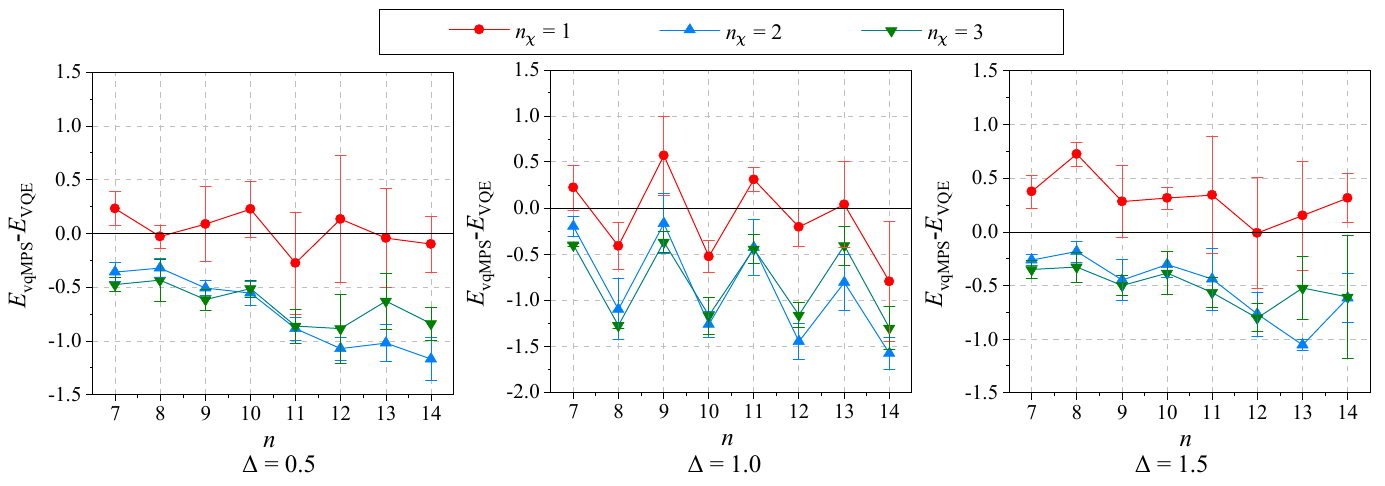}
  \caption{Results of numerical simulations for Heisenberg spin-chain Hamiltonian with $\Delta = 0.5$, $1.0$ and $1.5$, respectively. The base-line is the result of full-scale VQE, and $E_{\rm vqMPS}-E_{VQE}$ is the difference between the ground energy found from the sweeps of vqMPS and that of full-scale VQE. The vqMPS method reaches a comparable accuracy with VQE when $n_\chi = 1$, which requires only $2n_\chi + 1 = 3$ qubits per VQE process. Better accuracy can be achieved when $n_\chi = 2$ or $3$ because all the data are below the baseline. However, for some cases using $n_\chi=3$ can be less accurate compared with that of $n_\chi = 2$.}\label{fig:res}
\end{figure*}

We demonstrate the vqMPS method for finding the ground energy of Heisenberg XXZ spin-chain model
\begin{equation}\label{eq:XXZ}
  \hat{H}_{XXZ} = \sum_{i=1}^{N-1}[J(\hat{\sigma}_i^x\hat{\sigma}_{i+1}^x+\hat{\sigma}_i^y\hat{\sigma}_{i+1}^y)+\Delta\hat{\sigma}_i^z\hat{\sigma}_{i+1}^z]+h\sum_{i=1}^N\hat{\sigma}_i^z,
\end{equation}
where $\Delta$ is set to be $0.5$, $1.0$ and $1.5$, respectively. We test vqMPS method with bond dimensions fixed at $\chi = 2, 4$ or $8$, corresponding to using $n_\chi = 1, 2$ or $3$ ancillary qubits. In each case, the number of qubits involved within the optimization of a single site is $2n_\chi + 1$, and the qSVD or qreshape can reach fidelities exceeding 99\%. Note that when the full system is scaled up, the scale for optimizing each site remains the same.

We test our methods compared with the standard VQE method. The structure of the variational circuits are shown in Fig.~\ref{fig:qsvd}(c) the depth of the circuit is limited to $n$ for $n$-qubit VQE, and for the efficiency of simulation, we use only one $R_y$ gate on each qubit in each variational layer. The results are shown in Fig.~\ref{fig:res}. We performed 300 iterations and reached a good convergence for each optimization process. The maximum number of iterations is limited to control the computing cost. Since the depth of the circuits as well as the number of iterations are limited, the optimization process may not reach the ground truth. But in vqMPS, the VQE can reach a much higher ratio towards the optimal value owing to the reduced number of qubits. For VQE with less than 5 qubits, using such configurations can reach nearly optimal values, but for VQE with a large number of qubits, it requires both longer time for each step and more steps to converge. The vqMPS method can be scaled up to tens of qubits, and the time complexity is approximately linear. For all the cases, we find the vqMPS can reach a comparable accuracy compared with VQE when using merely $n_\chi = 1$ qubit for connecting (i.e. $\chi = 2$), namely each local optimization by VQE involves at most 3 qubits. The accuracy can be further improved for $n_\chi = 2$ and $3$, using 5 or 7-qubit VQE. For some cases using $n_\chi=3$ can be less accurate compared with that of $n_\chi = 2$, again reflecting the difficulty in large-scale VQE.

\section{Conclusion}\label{sec:summary}

In this paper, we propose the qMPS model and the variational qMPS method, which is actually the quantum version of the DMRG method. We update the quantum methods for preparing a qMPS in the canonical form. The whole method can be done on quantum devices and we demonstrate this method through numerical simulation. The state is updated locally, and therefore can greatly reduce the number of qubits involved in variational algorithms, while maintaining the accuracy. For qMPS, the auxiliary dimensions are represented by qubits, and can bring in logarithmic space complexity in storing tensor data. However, the result is strongly correlated with the quality of the optimization, including the procedures of QSVD and the VQE. It is worthy pointing out that for state with concentrated spectrum, the fidelity of the recovered state from the decomposed results can be nearly identity, but the base vectors learned corresponding to small singular values can be biased. The accuracy of vqMPS can be further improved using deterministic algorithms on quantum devices with error corrections. These models and methods show the possibilities of transplanting efficient classical algorithms to quantum devices, as well as the means for quantum-classical hybrid tasks.

\begin{acknowledgments}

We appreciate the helpful discussion with other members of the QUANTA group. This work was funded by the National Key R\&D Program of China under Grant No. 2024YFB4504001, the National Natural Science Foundation of China under Grant No. 62401572, the Aid Program for Science and Technology Innovative Research Team in Higher Educational Institutions of Hunan Province, and the Open Research Fund from State Key Laboratory of High Performance Computing of China under Grant No. 202401-06.

\end{acknowledgments}

\bibliographystyle{unsrt}

\end{document}